\documentclass[nohyper,12pt,letterpaper]{JHEP3}
\usepackage{amsfonts,amssymb}
\newcommand{\be}{\begin{equation}}
\newcommand{\ee}{\end{equation}}
\newcommand{\ben}{\begin{displaymath}}
\newcommand{\een}{\end{displaymath}}
\newcommand{\bea}{\begin{eqnarray}}
\newcommand{\eea}{\end{eqnarray}}
\newcommand{\bean}{\begin{eqnarray*}}
\newcommand{\eean}{\end{eqnarray*}}

\def\ap {\alpha'}

\newcommand{\calp}{\mbox{${\cal P}$}}

\newcommand{\bbr}[1]{\mbox{${\mathbb R}^{#1}$}}
\newcommand{\ads}[1]{\mbox{${AdS}_{#1}$}}
\newcommand{\adss}[2]{\mbox{$AdS_{#1}\times {S}^{#2}$}}

\newcommand{\bra}[1]{\mbox{$\langle #1 |$}}
\newcommand{\ket}[1]{\mbox{$| #1 \rangle$}}

\newcommand{\eg}{{\it e.g.}}
\newcommand{\ie}{{\it i.e.}}

\newcommand{\tr}{\mbox{Tr}}

\newcommand{\beq}{\begin{equation}}
\newcommand{\eeq}{\end{equation}}
\newcommand{\beqr}{\begin{displaymath}}
\newcommand{\eeqr}{\end{displaymath}}
\newcommand{\beqa}{\begin{eqnarray}}
\newcommand{\eeqa}{\end{eqnarray}}
\newcommand{\beqar}{\begin{eqnarray*}}
\newcommand{\eeqar}{\end{eqnarray*}}
\newcommand{\cH}{{\cal H}}
\newcommand{\cN}{{\cal N}}
\newcommand{\cD}{{\cal D}}
\newcommand{\cO}{{\cal O}}

\newcommand{\cA}{{\cal A}}

\newcommand{\half}{\ensuremath{\frac{1}{2}}}

\newcommand{\N}[1]{\ensuremath{\cN=#1}}

\newcommand{\vn}{\ensuremath{\vec{n}}}
\newcommand{\tl}{\ensuremath{\tilde{\lambda}}}
\newcommand{\ct}{\ensuremath{\cos\theta}}
\newcommand{\st}{\ensuremath{\sin\theta}}
\newcommand{\ps}{\ensuremath{\partial_\sigma}}

\newcommand{\pp}{\ensuremath{\dot{\varphi}}}
\newcommand{\vphi}{\ensuremath{\varphi}}
\newcommand{\pt}{\ensuremath{\partial_t}}
\newcommand{\pta}{\ensuremath{\partial_{\tau}}}
\newcommand{\ts}{\ensuremath{\tilde{\sigma}}}

\title{\LARGE Spin chains and string theory}

\author{Martin Kruczenski \\
        Department of Physics, Brandeis University \\
        Waltham, MA 02454.

E-mail: \email{martink@brandeis.edu}}

\abstract{Recently, an impressive agreement was found between anomalous dimensions of certain operators in \N{4} SYM 
and rotating strings with two angular momenta in the bulk of \adss{5}{5}. A one-loop field theory computation, 
which involves solving a Heisenberg chain by means of the Bethe ansatz agrees with the large angular 
momentum limit of a rotating string. We point out that the Heisenberg chain can be equally
well solved by using a sigma model approach. Moreover we also show that a certain limit, akin to the BMN limit, 
leads exactly to the same sigma model for a string rotating with large angular momentum. The agreement is then 
at the level of the action. As an upshot we propose that the rotating string should be identified with a coherent, 
semi-classical state built out of the eigenstates of the spin chain. The agreement is then complete. For example 
we show that the mean value of the spin $\langle \vec{S} \rangle$ gives, precisely, the position of the string in 
the bulk. This suggests a more precise formulation of the AdS/CFT correspondence in the large-N limit and
also indicates a way to obtain string theory duals of other gauge theories.
}


\preprint{\tt{BRX TH-539} \\
          \tt{hep-th/0311203}  }

\begin{document}

%%%% INTRODUCTION
\section{Introduction}

Maldacena's AdS/CFT correspondence~\cite{malda} promised to reveal the deep relation between gauge theory and
string theory that was conjectured to exist in 't Hooft's large-N limit~\cite{largeN}. In a recent paper Berenstein, 
Maldacena and Nastase~\cite{Berenstein:2002jq} made the first steps in that direction by showing
that certain operators in the boundary theory corresponded to string excitation in the bulk. After
that, it was observed that such relation followed from a more general relation between that of semi-classical rotating 
strings in the bulk~\cite{Gubser:2002tv} and certain operators in the boundary. A lot of activity followed those papers. 
In particular many new rotating solutions were found\footnote{There is a large body of literature in the subject. For the purpose 
of this paper the relevant reference is~\cite{Frolov:2003xy}. The reader is urged to consult the recent 
review~\cite{Tseytlin:2003ii} for a nice introduction to the subject and a complete set of references.}. In a parallel 
development, Minahan and Zarembo~\cite{Minahan:2002ve} observed that the one-loop anomalous dimension of operators composed 
of scalars in \N{4} SYM theory follows from solving and integrable spin chain\footnote{In QCD the relation between spin chains 
and anomalous dimensions had already been noted in~\cite{Braun:1998id}.}. Subsequently~\cite{Beisert:2003xu}, much attention 
was devoted to a subset of operators given by
\beq
\cO^{J_1,J_2} = \tr  ZZXZXX\ldots ZX,
\label{Oj1j2}
\eeq
where the right hand side contains and arbitrary permutation containing a number $J_1$ of Xs and $J_2$ of Zs. Here we denote
$X=\Phi^1+i\Phi^2$, $Z=\Phi^3+i\Phi^4$ and $\Phi^a$, $a=1\ldots 6$ are the adjoint scalars of \N{4}. There are
as many such operators as different  permutations of the $X$ and $Z$ one can make up to cyclic permutations.
In the free theory all these operators have conformal dimension $\Delta_0=J_1+J_2=J$. The one loop anomalous dimension
can be obtained from the 1-loop dilatation operator which, acting on these operators, 
takes the form\footnote{All we say in this paper is valid (and is relevant) only in the large-N limit.}~\cite{Beisert:2003tq}:
\beq
D_{\mbox{1-loop}} = \tl \sum_1^J (\frac{1}{4} - \vec{S}_i \vec{S}_{i+1}),\ \ \mbox{with}\ \ 
\tl = \frac{\lambda}{4\pi^2}=\frac{g_{YM}^2 N}{4\pi^2} .
\label{D1l}
\eeq
To apply $D_{\mbox{1-loop}}$ to the $\cO^{J_1,J_2}$, one should consider $\cO^{J_1,J_2}$ as a spin $1/2$ chain 
identifying \eg\ $X$ with a spin down state and $Z$ with a spin up:
\beq
ZZXZXX\ldots ZX \ \ \ \ \Longleftrightarrow \ \ \ \ | \uparrow\ \uparrow\ \downarrow\ \uparrow\ \downarrow\ \downarrow
                                                      \ \uparrow\ \downarrow\ \rangle.
\label{spins}
\eeq
After this identification the spin operators $\vec{S}$ act in the usual manner. The only point is that the trace in (\ref{Oj1j2})
implies that we have to consider periodic chains ($\vec{S}_{J+1}=\vec{S}_1$) and zero momentum states, \ie\ invariant under
cyclic permutations. With all these in place, the computation of 1-loop anomalous dimensions reduces to the diagonalization
of a spin $\half$ ferromagnetic Heisenberg chain. The length of the chain\footnote{Note that this differs from the conventions 
used in the condensed matter literature where the strength of the interaction is usually denoted as $J$ and the length of 
the chain as $N$ or $L$.} is $J$ and the coupling constant is $\tl$.  The spin chain can be diagonalized by using the 
Bethe ansatz (for the case in hand see \cite{Minahan:2002ve,Beisert:2003xu}). This allowed the 
authors of~\cite{Beisert:2003xu} to obtain an operator, 
linear combination of the $\cO^{J_1,J_2}$, with anomalous dimension given by
\beq
\gamma = \frac{2\lambda}{\pi^2 J} K(x) (E(x)-(1-x)K(x)), \ \ \ \ \frac{J_2}{J}= 1-\frac{E(x)}{K(x)},
\label{gamma1}
\eeq
where $x$ is an auxiliary parameter that should be eliminated using the second equation. The functions $E(x)$ and
$K(x)$ are standard elliptic integrals. This gives the anomalous dimension $\gamma$ as a function 
of $J=J_1+J_2$ and $J_2$. Although this is in itself a remarkable achievement, the most amazing thing is that, as shown 
in~\cite{Beisert:2003ea}, it precisely agrees with a string computation done in the dual \adss{5}{5} theory. The string 
calculation involves finding a rotating string with the same quantum numbers and identifying its energy with the conformal 
dimension $E=J+\gamma$ as in~\cite{Gubser:2002tv}. The result is~\cite{Frolov:2003xy}:
\beqa
\left(\frac{J+\gamma}{K(x)}\right)^2 - \left(\frac{J_1}{E(x)}\right)^2 = \frac{4\lambda}{\pi^2} x ,\ \ \ \  
\left(\frac{J_2}{K(x)-E(x)}\right)^2 - \left(\frac{J_1}{E(x)}\right)^2 = \frac{4\lambda}{\pi^2}.
\label{gamma2}
\eeqa
This does not look the same as eq.(\ref{gamma1}). However we should remember that (\ref{gamma1}) is valid only at first 
order in $\lambda$. Interestingly, (\ref{gamma2}) also has an expansion in powers of $\lambda$ (more 
precisely $\lambda/J^2$) 
and the first term agrees exactly with (\ref{gamma1}). This is the main result of~\cite{Beisert:2003ea}.
Subsequently, a more detailed agreement between the integrable structures\footnote{See \eg\ 
\cite{Arutyunov:2003za,Arutyunov:2003uj} for an account of the relation between integrable systems and 
rotating strings.} of the classical string and the spin chain 
was found for these solutions in \cite{Arutyunov:2003rg}.  

In this paper we go one step further and show that the spin chain system, in the limit of large number of sites, 
is described by a sigma model which precisely agrees with the sigma model obtained from the rotating string in the 
relevant limit. From the Heisenberg chain side the result is well-known~\cite{Fradkin}, the low energy modes
can be described through an effective action which, at the order in $1/J$ we need, has only one undetermined 
coefficient that can be fixed from the BMN limit. From the string theory side it amounts to taking a limit of 
the Polyakov action. The identification then goes beyond a particular solution. 
Moreover this gives a precise mapping of the states. For example we show that the mean value of the spin at a given site 
is the same as the position of the string in the bulk. That is, given a state of the spin chain we can map it into a 
string configuration making the identification complete. In the states we find, the spins are precessing with 
exactly the same motion as the string. 

The results of \cite{Beisert:2003ea,Arutyunov:2003rg} are a manifestation of the agreement between 
these two ways of describing a spin chain, 
in terms of Bethe eigenstates or through a low energy effective action. The AdS/CFT correspondence, on the other hand,
states that one of those descriptions also corresponds to a string rotating in $AdS_5\times S_5$ 
(in a particular limit). 

In the reminder of the paper, we analyze briefly the higher loop dilatation operator which leads us to suggest 
that a more precise formulation of the large-N Maldacena conjecture is that the classical action for string theory 
in global \adss{5}{5} is equivalent to the long wave length limit of the dilatation operator acting on single trace 
operators. By long wave length limit we mean when taking traces of a large number of operators. Of course this is a 
well known idea since we are just saying that the dilatation operator is the hamiltonian in global coordinates. Perhaps 
the new ingredient, if any, is that we are suggesting a slightly more precise way to extract the string action from the 
field theory. Unfortunately, it is not as precise as we would like it to be since, as we shall 
see, the exact way of mapping the sigma models at higher loops is not completely clear to us.
Even knowing the exact mapping, in practice it is not feasible to do an all loop computation in the field theory, but we 
believe it is interesting that there is a way, at least in principle to embed the string theory in the field theory. Here 
we do just that at 1-loop by using previously known results. Finally, it is clear that this type of calculation can be 
extended to any gauge theory suggesting a way to obtain a string dual of a gauge theory.  By constructing the corresponding 
spin chain and taking the long wave length limit, a sigma model will emerge that describes a string moving in some 
background. With some effort (\ie\ computing more loops) or some guess-work one might be able to reproduce the 
whole background. The known alternative method is to construct the theory using D-branes and then finding the 
supergravity background.

%%%%%SECTION 2
\section{Heisenberg chain and ferromagnetic sigma model}
\label{Heisen}

The ferromagnetic spin $s$ Heisenberg chain is exactly equivalent to a discrete sigma model known as the ferromagnetic
sigma model. By discrete sigma model we mean that there are a discrete number of sites with continuum variables
leaving on them. In the long wave length limit this reduces to a usual sigma model\footnote{Some times the word sigma model 
implies a particular form of the action. Here we just mean a field theory with a target space given by a compact manifold, in 
this case the coset space $SU(2)/U(1)=S^2$.} up to corrections $1/J$ where $J$ is the length of the chain.

Since this is well known, in this section we just sketch where this sigma model comes from and its main properties. 
We refer the reader to \eg~\cite{Fradkin} for a detailed account and references to the original work.
To obtain a sigma model one defines a set of coherent states labeled by two angles $\phi$ and $\theta$ and defined as
\beq
\ket{\vn} = e^{iS_z\phi} e^{iS_y\theta} \ket{ss},
\label{nstate}
\eeq
where $\vn$ is a unit vector with components 
\beq
\vn=(\sin\theta \cos\phi,\sin\theta\sin\phi,\cos\theta),
\label{npara}
\eeq
and $\ket{ss}$ is a state of total spin $s$ and maximal $z$-projection $S_z=s$.
The identity can be written as
\beq
1=\left(\frac{2s+1}{4\pi}\right)\,\int \, \sin^2\theta d\theta d\phi\, \ket{\vn} \bra{\vn}.
\label{identity}
\eeq
Inserting this in the standard derivation of the path integral from the Hamiltonian approach one finds that a transition
amplitude can be computed in terms of a path integral 
\beq
\cA = \int \cD \vn e^{iS(\vn)},
\label{propagator}
\eeq
with action
\beq
S (\vn) = s \sum_k \int dt \int_0^1 d\tau \vn_k.(\partial_t n_k \times \partial_\tau \vn_k) 
          - \frac{\tl s^2}{2} \int dt \sum_k  [\vn_k-\vn_{k+1}]^2.
\label{discreteaction}
\eeq
The first term is a Wess-Zumino term proportional to the area spanned between the trajectory and the North pole.  
Its definition requires the introduction of an additional coordinate $\tau$ with $\theta(\tau=1)=0$. The variation of the
Wess-Zumino term is given just by a boundary term at $\tau=0$ which is the only way in which it enters in the equations of motion. 
The important point to note is that this is exactly equivalent to the Heisenberg chain, no approximation was made. At first
sight this does not seem possible since now we have, in appearance, an infinite number of states per site. However this is not
so. The Wess Zumino term is equivalent to a magnetic charge $s$ at the center of the $S^2$ over which $\vn$ is moving. Therefore,
one should find the Landau levels. For charge $s$ one finds precisely $2s+1$ levels. The gap to the other levels is infinite since
no kinetic term $\partial_t\vn\partial_t\vn$ is present. Another consequence is that, since the action is 
linear in time derivatives, near the ground state the dispersion relation is quadratic ($\omega\sim k^2$) as is know to be for
ferromagnetic magnons\footnote{This in not the case for antiferromagnets were $\omega\sim k$ and which are described
by the sine-Gordon model.}. In our case this is the BMN limit of operators (\ref{Oj1j2}) where $\gamma \sim n^2$. 

 As it is, there is no advantage in using the sigma model. The interest arises when one takes the limit of very large chains,
or large $J$ in operator language. In that limit it makes sense to consider the long wave length limit of (\ref{discreteaction}) 
and consider a continuous coordinate $\sigma$ running from $0$ to $J$. The action is
\beq
S (\vn) = s \int dt d\sigma \int_0^1 d\tau \sin\theta\left(\partial_\tau\phi \pt\theta - \pt\phi\partial_\tau\theta\right)
-\frac{\tl s^2}{2} \int d\sigma dt\left( (\partial_\sigma\theta)^2+\sin^2\theta (\ps\phi)^2\right),
\label{contaction}
\eeq
where we used the parameterization (\ref{npara}).
 From here we obtain the momenta, Hamiltonian and momentum density in direction $\sigma$:
\beqa
\calp_{\phi} &=& S_z = -s\int d\sigma \int_0^1 d\tau \sin\theta \partial_\tau\theta =-s \int\cos\theta\,d\sigma, \\
\calp_{\theta} &=& s \int d\sigma \int_0^1 d\tau \sin\theta \partial_\tau\phi, \\
\cH &=& \calp_{\phi} \pt\phi + \calp_{\theta} \pt\theta - L = \frac{\tl s^2}{2} \int d\sigma  
\left( (\partial_\sigma\theta)^2+\sin^2\theta (\ps\phi)^2\right), \\
P &=& \int T_{01} d\sigma = -s\int_0^J\cos\theta \partial_\sigma \phi = 0, \label{P=0}
\label{momenta}
\eeqa
where the last equality is a condition on the solutions. Form the operator point of view this means that we consider
operators invariant under cyclic permutations (see discussion below eq.(\ref{spins})). 
The ground state of this system is with all spins aligned parallel since it is ferromagnetic. On the other hand, we 
are interested here in highly excited states. Since quantum numbers are large we expect that they will be well described by classical 
solutions of the equations of motion. This classical solutions should emerged in quantum theory as superposition of a large
number of eigenstates of the same approximate energy. Since this argument cannot be applied to individual spins because they
have only two states, what we consider is that $\vn$ describes the average spin over long domains that form since
the ferromagnetic Heisenberg chain has an ordered ground state. Namely the spins tend to be parallel at short distances
with fluctuations occurring only at long distances. 

We should also comment that in taking the continuum limit the coupling constants can be renormalized. The number in front 
of the Wess-Zumino term has to be a half integer by a topological argument so it cannot be renormalized but the constant 
in front of the gradient in principle can. That such renormalization does not occur can be checked by computing the 
spectrum of magnons near the ground state and comparing with the known exact result\footnote{This comment refers to the
Heisenberg chain, not the whole field theory.}. In our case one would say that one checks that there is no renormalization 
of the coupling constant by comparing with the BMN limit. 

In principle this is all we wanted to do in this section. In the next one we show that exactly the same sigma model 
follows from taking the large angular momentum limit of a string rotating in $S^3$ (which is the relevant part of \adss{5}{5}).
However the reader will probably feel much better if we use the sigma model to reproduce for example (\ref{gamma1})
showing that it actually gives the same result as the Bethe ansatz.

Before doing that however, we want to argue that since we are only going to look for classical solutions we can integrate
by parts the Wess Zumino term and convert it into a more conventional term. This amounts to write the volume form 
$\omega_{[2]}=\sin\theta d\theta d\phi$  of the sphere as an exact form $\omega_{[2]}=dA$, $A=-\cos\theta d\phi$. This is 
singular at the poles but for studying the classical solutions we can ignore that fact. The Lagrangian becomes
\beq
S (\vn) = - s \int d\sigma dt \cos\theta \partial_t\phi 
-\frac{\tl s^2}{2} \int d\sigma dt\left( (\partial_\sigma\theta)^2+\sin^2\theta (\ps\phi)^2\right).   
\label{actionfinal}
\eeq

\subsection{Particular solutions}

We want to find now a particular classical solution with given angular momentum. Note that $J$ is already given by
the length of the chain. On the other hand, we can fix $S_z$, the total component of the spin in direction $z$. 
For the operators $\cO^{J_1,J_2}$ we have $S_z = (J_2-J_1)/2$. The equations of motion that follow from the Lagrangian are
\beqa
\st \pt\theta + \tl s \ps\left(\sin^2\theta \ps \phi\right) &=&0, \\
\st \pt \phi + \tl s \partial_\sigma^2 \theta - \tl s \st \ct (\ps \phi)^2 &=& 0.
\label{eqom}
\eeqa
The boundary conditions are
\beq
\phi(\sigma=J,t)=\phi(\sigma=0,t), \ \ \ \theta(\sigma=J,t)=\theta(\sigma=0,t),
\label{bc}
\eeq
and the zero momentum condition (\ref{P=0}) is still 
\beq
P_{\sigma} = -\half\int \cos\theta \partial_\sigma \phi =0 .
\label{P=0b}
\eeq
Now we make the ansatz $\partial_\sigma\phi=0$ which immediately satisfies this last 
condition\footnote{It would be interesting to study solutions with $\partial_\sigma\phi\neq 0$ since (\ref{P=0b})
does not necessarily imply $\partial_\sigma\phi=0$. This other solutions should also agree with supergravity.}. The first 
equation immediately implies that $\pt\theta=0$. The second 
equation then implies that $\pt^2\phi=0$. So we put $\pt\phi=w$. Finally the only equation that we have to solve is
\beq
\ps^2\theta + \frac{w}{\tl s} \st =0.
\label{eqom2}
\eeq
This integrates to
\beq
\ps\theta = \pm \sqrt{a+b\cos\theta}, \ \ \ \mbox{with} \ \ \ b=\frac{2w}{\tl s},
\label{eqom3}
\eeq
and $a$ is a constant of integration. Two possibilities arise. If $a>|b|$ then the square root never vanishes. This means
that $\theta$ keeps increasing and then the periodic condition should read $\theta(J) = \theta(0) + 2\pi n$, for some integer $n$.
The other possibility is that $b>|a|$. Then, at $\theta_0=\arccos(-a/b)$ the square root becomes zero and we can change branches,
namely the ``particle'' returns oscillating between $-\theta_0<\theta<\theta_0$. In this paper we concentrate in this solution and 
leave the other case for the interested reader. 
 
With this solution we compute the energy and angular momentum as:
\beqa
S_z &=& -4s\int_0^{\theta_0} \frac{\cos\theta}{\sqrt{a+b\ct}} d\theta ,\\
J &=& \int_0^J d\sigma = 4 \int_0^{\theta_0} \frac{d\theta}{\sqrt{a+b\ct}} ,\\
\gamma &=& E = 2\tl s^2 \int_0^{\theta_0} \frac{a+b\ct}{\sqrt{a+b\ct}} d\theta = \frac{\tl s^2}{2}\left(a J-\frac{b}{s}S_z\right).
\label{constants}
\eeqa
The integrals can be performed in terms of elliptic integrals~\cite{Gradshteyn}\footnote{For simplicity of comparison we follow the 
notation of~\cite{Beisert:2003ea} rather than that in~\cite{Gradshteyn} which differ in the argument of the elliptic integral
being $x$ or $\sqrt{x}$} 
giving
\beqa
S_z &=& -4s\sqrt{\frac{2}{b}}\left\{2E(x)-K(x)\right\} ,\\
J &=& 4 \sqrt{\frac{2}{b}} K(x) ,\\
 x &=& \frac{a+b}{2b}.
\label{EK}
\eeqa
Finally, using that $S_z = \frac{J_2-J_1}{2}$, simple algebra leads to 
\beqa
\frac{J_2}{J} &=& (s+\half)-2s\frac{E(x)}{K(x)}, \\
\gamma &=& E = \frac{\tl}{J} 32 s^2 K(x)\left[E(x)-(1-x)K(x)\right].
\label{result1}
\eeqa
 If we now replace $s=\frac{1}{2}$ and, from (\ref{D1l}), $\tl \rightarrow \lambda/4\pi^2$ we get 
precisely (\ref{gamma1}).
%\footnote{It is not known to me if this agreement between the Bethe ansatz and the sigma model
%has been checked before.}.

 What we have just obtained is a sigma model that reproduces the Bethe ansatz result. But, actually, that is already done  
by the string theory sigma model, albeit in a particular limit. It is obvious that both should be related. In the next section
we derive this same sigma model directly from the string theory action. 

 Before doing so, let us analyze the solution we have just obtained a little further. At an instant in time the position of
the spins is given by $\theta(\sigma)$. We see that, as sigma is varied, the end point of $\vec{n}$ goes from $\theta=0$
to $\theta=\theta_0$ and back. Then to the other side. This looks like a folded string already. Furthermore, each point is precessing
around the $z$ axis with the same angular velocity $w$. So the configuration looks exactly like a rotating string!. The only
thing lacking from the Frolov-Tseytlin solution is the motion of the center of mass. We will see 
that to get this sigma model from the string picture one has to subtract the motion of the center of mass in a similar way as is done 
in the pp-wave limit \footnote{Actually, in our case we have to subtract also another ``fast'' rotation as
it will become apparent in the next section. I am grateful to A. Tseytlin for pointing this out.}.
 A final caveat is that when we talk about time dependent solutions, time is the conjugate variable to the Hamiltonian which here
is actually the dilatation operator. So, from the field theory point of view  we are thinking of an operator with some 
weird properties under scaling. This makes sense however when we think of the state/operator correspondence which 
exists in conformal theories. Under that mapping, such an operator maps into a time dependent state which is precisely what we want 
since the string is a time dependent state. That the relation between the operator and the string includes a state/operator 
correspondence is well known from the rules of the AdS/CFT correspondence.

%%%%%SECTION 3
\section{Rotating string}
\label{Rotating}

The rotating string solutions corresponding to the operators discussed in the previous section were found in~\cite{Frolov:2003xy}. The
relevant part of the metric in the coordinates used in that paper is
\beq
ds^2 = -dt^2 + d\Omega_{[3]}^2 = -dt^2 + d\psi^2 + \cos^2\psi d\phi_1^2 + \sin^2\psi d\phi_2^2.
\label{metric}
\eeq
Changing coordinates to $\phi_1 = \varphi_1 + \varphi_2$, $\phi_2 = \varphi_1-\varphi_2$ we get a metric
\beq
ds^2 = -dt^2 + d\psi^2 + d\varphi_1^2 + d\varphi_2^2 + 2\cos(2\psi) d\varphi_1 d\varphi_2.
\label{metric2}
\eeq
In the Frolov-Tseytlin solution the center of mass moves with angular velocity $w$ in direction $\varphi_1$. In the limit
of large angular momentum $w$ tends to one. It makes sense then to do a change of coordinates
\beq
\varphi_1 \rightarrow t + \varphi_1 ,
\label{chcoord}
\eeq
and the metric becomes 
\beq
ds^2  = 2 dt d\varphi_1 + d\psi^2 + d\varphi_1^2 + d\varphi_2^2 + 2\cos(2\psi) dt d\varphi_2 + 2\cos(2\psi) d\varphi_1d\varphi_2.
\label{metric3}
\eeq
After making the gauge choice $t=\kappa\tau$, the Polyakov action describing a string in this background becomes:
\beqa
S &=& \frac{R^2}{4\pi\ap} \int G_{\mu\nu} \pta X^\mu \pta X^\nu - G_{\mu\nu} \ps X^\mu \ps X^{\nu} \\
  &=  \frac{R^2}{4\pi\ap} \int & 2\kappa \pp_1 + \dot{\psi}^2 + \pp_1^2 +\pp_2^2 + 2\cos(2\psi) \kappa \pp_2 + 2\cos(2\psi) \pp_1\pp_2 \\
  &&  - \psi'^2 - \vphi_1'^2 - \vphi_2'^2 - 2\cos(2\psi)\vphi_1'\vphi_2' ,
\eeqa
where we denote derivatives with respect to $\tau$ with a dot and those with respect to $\sigma$ with a prime.
The Virasoro constraints are:
\beqa
G_{\mu\nu} \pta X^\mu \ps X_\nu &=& 2\kappa \vphi'_1 + \dot{\psi} \psi' + \pp_1 \vphi'_1 + \pp_2 \vphi'_2 
  + 2\cos(2\psi) \kappa \vphi'_2  + 2\cos(2\psi) \pp_1 \vphi'_2 + \nonumber \\
 && + 2\cos(2\psi) \pp_2 \vphi'_1 =0 ,
\eeqa
and
\beqa
G_{\mu\nu} \pta X^\mu \pta X^\nu + G_{\mu\nu} \ps X^\mu \ps X^\nu  &=& 2\kappa \pp_1 + \dot{\psi}^2 
      + \pp_1^2 +\pp_2^2 + 2\cos(2\psi) \kappa \pp_2 + 2\cos(2\psi) \pp_1\pp_2 \nonumber\\
  &&  + \psi'^2 + \vphi_1'^2 + \vphi_2'^2 + 2\cos(2\psi)\vphi_1'\vphi_2' =0, 
\eeqa
where we have also used the gauge choice $t=\kappa\tau$.

Up to now we did not do any approximation. Now we are going to assume that the motion of the string is mainly captured by the rotation
we just did through the change of coordinates. Specifically, we assume that all time derivatives are small. If we drop time derivatives
altogether then the string will be moving at the speed of light and the only solution is the BPS state corresponding to a massless
string moving around the circle. Looking at the action and the constraints one sees that a non-trivial limit is obtained
by taking
\beq
 \dot{X}^\mu \rightarrow 0, \ \ \ \kappa\rightarrow \infty, \ \ \ \mbox{with}\ \ \ \kappa\dot{X}^{\mu} \ \ \mbox{fixed.} 
\label{limit}
\eeq
Here $X^{\mu}$ denotes any coordinate except $t$. Later on, we will see that taking this limit in the Frolov-Tseytlin solution
precisely corresponds to the large angular momentum limit with $J \sim \kappa$. 

As a comment we want to indicate that this limit, although similar to the BMN limit\footnote{It also has some similarities
with the so called wrapped or non-relativistic limit~\cite{wrapped}. For the present case some related 
ideas appeared in~\cite{Gorsky:2003nq} and \cite{Mikhailov:2003gq}.}~\cite{Berenstein:2002jq} is somewhat 
different. Here, we are not 
taking the limit directly 
in the metric but in the action. As a result we keep terms such as $\vphi_2'^2$ which would not be be there if we were 
zooming into a geodesic. 

In the limit (\ref{limit}), the action reduces to 
\beq
S = \frac{R^2}{4\pi\ap} \int 2\kappa \pp_1 + 2\cos(2\psi) \kappa \pp_2 - \psi'^2 - \vphi_1'^2 - \vphi_2'^2 
- 2\cos(2\psi)\vphi_1'\vphi_2' ,
\eeq
and the constraints become
\beqa
&& 2\kappa \vphi'_1 +  2\cos(2\psi) \kappa \vphi'_2 =0, \\
&& 2\kappa \pp_1 + 2\cos(2\psi) \kappa \pp_2 + \psi'^2 + \vphi_1'^2 + \vphi_2'^2 + 2\cos(2\psi)\vphi_1'\vphi_2' = 0.
\eeqa
These constraints determine $\vphi_1$ as a function of $\sigma$ and $\tau$. Since from the action there is no 
equation of motion for $\vphi_1$ we can always satisfy them. The only caveat is that $\vphi_1$ is determined twice.
However, one can check using the equations of motion that $\dot{\vphi_1}'$ as determined by the first or second constraint agree.
Also since the string is closed, $\vphi_1$ is a periodic function of sigma which implies
\beq
0 = \int d\sigma \vphi_1' = - \int d\sigma \cos(2\psi) \vphi_2',
\eeq
which, as we will see below is the condition (\ref{P=0}). 

Replacing the first constraint in the action we get
\beq
S = \frac{R^2}{4\pi\ap} \int 2\kappa \pp_1 + 2\cos(2\psi) \kappa \pp_2 - \psi'^2 - \sin^2(2\psi) \vphi_2'^2 ,
\eeq
which already looks quite similar to the sigma model we had before. To make the agreement precise we first compute the
angular momentum
\beq
J = \calp_{\vphi_1} = \frac{R^2}{4\pi\ap} 2\kappa \int_0^{2\pi} d\sigma = \int_0^J d\ts,
\eeq
where we defined
\beq
\ts = \frac{R^2}{4\pi\ap}2\kappa\sigma,
\eeq
so that the length of the chain is $J$ as before. To compare the energies we rescale $\tau$ into $t=\kappa \tau$. Also
one can see that the angles are related by
\beq
\vphi_2 = -\frac{1}{2} \phi, \ \ \ \ \psi=\frac{1}{2} \theta,
\label{angles}
\eeq
which is related to the fact (at least for $\vphi$) that the chain has sites with spin $1/2$. Finally we use the 
AdS/CFT relation $R^2/\ap=\sqrt{\lambda}$ to get the action
\beq
S = -\frac{1}{2} \int dtd\ts \cos\theta \pt \phi - \frac{\lambda}{32\pi^2}\int dt d\ts 
                             \left((\ps\theta)^2 + \sin^2\theta (\ps\phi)^2\right),
\eeq
which precisely agrees with (\ref{actionfinal}) after we put $s=1/2$. Also the identification of the angles (\ref{angles})
implies that we can map directly a configuration $\vn(\sigma)$ into a particular shape of the string in the bulk regardless
if it is a solution or not. Since $\vn$ is the average value of the spin at a site, $\bra{\vn} \vec{S} \ket{\vn} =s \vn$
one can identify the average value of the spin with the position of the corresponding portion of the string in the bulk. 
This is slightly more than we are entitled to argue since the calculations are valid in the long
wave length limit only. However it is a natural conjecture to make. It is not trivial because
it implies that if we look at the string carefully enough we will see the discrete nature of the spin chain. This can be 
a generic feature of string theory or just a peculiarity of the limit we are taking. 

\subsection{Particular solutions}
It is clear that, since the actions agree, we get the same solutions. However we still need to check that at
least in the case of the Frolov-Tseytlin solution the limit we have taken agrees with the one taken in~\cite{Beisert:2003ea}.
The solution for the rotating string in \cite{Frolov:2003xy} follows from the ansatz:
\beq
t=\kappa\tau, \ \ \ \phi_1=w_1\tau, \ \ \ \phi_2=w_2\tau, \ \ \ \psi=\psi(\sigma). 
\eeq
The relations between the energy, angular momentum and angular velocities can be cast in the form \cite{Beisert:2003ea}:
\beqa
x &\equiv& \frac{\kappa^2-w_1^2}{w_2^2-w_1^2}, \\
E &=& \kappa ,\\
1 &=& \frac{1}{w_1} J_1 + \frac{1}{w_2} J_2, \\
J_1 &=& \frac{2w_1}{\pi\sqrt{w_2^2-w_1^2}} E(x), \\
\frac{2}{\pi} K(x) &=& \sqrt{w_2^2-w_1^2},
\eeqa
where we remind the reader that $E$, $J_1$ and $J_2$ are rescaled by $\sqrt{\lambda}$ in these equations. 
The limit we described before corresponds to taking
\beq
w_1=\kappa - \epsilon_1 , \ \ w_2=\kappa+\epsilon_2, \ \ \ \epsilon_{1,2}\rightarrow 0, \ \ \kappa\rightarrow\infty,\ \ 
\mbox{with}\ \ \kappa\epsilon_{1,2} \ \ \mbox{fixed}.
\eeq
For the moment however we just assume that $\epsilon_{1,2}\rightarrow 0$. In this limit we get
\beqa
J_1 &=& \frac{1}{\pi}\sqrt{\frac{2\kappa}{\epsilon}} E(x_0), \\
J_2 &=& \kappa - \frac{1}{\pi} \sqrt{\frac{2\kappa}{\epsilon}} E(x_0),
\eeqa
 where $\epsilon=\epsilon_1+\epsilon_2$ and $x_0=1/(1-\epsilon_2/\epsilon_1)$ is the limiting value of $x$. This implies
\beq
\frac{J_2}{J_1} = -1 + \frac{\pi}{\sqrt{2}} \frac{\sqrt{\kappa\epsilon}}{E(x_0)}.
\eeq
We see that in this limit both angular momenta diverge. However, if we want the ratio $J_2/J_1$ to be fixed at an arbitrary
value, we need to consider $\kappa\rightarrow\infty$ with $\kappa\epsilon$ fixed. It is also easy to show that at lowest order
we have
\beq
E=J=\kappa,
\eeq
so $\kappa\rightarrow\infty$ precisely corresponds to $J\rightarrow\infty$.

%%%%%SECTION 2
\section{Higher loops}
\label{loops}

 In this section we begin the study of the sigma model for higher loops in the gauge theory side. It was shown 
in~\cite{Beisert:2003tq} that the two-loops dilatation operator has the form\footnote{In~\cite{Beisert:2003tq}
it was actually written in terms of permutation operators. The form given here is equivalent.}  
\beq
D_{\mbox{2-loops}} = \frac{\lambda^2}{128\pi^4} \left[ -\frac{3}{2} J +8\sum_k \vec{S}_k \vec{S}_{k+1}
                     -2 \sum_k \vec{S}_{k} \vec{S}_{k+2}\right] .
\label{2loops}
\eeq
This is a spin chain with first and second neighbors interactions. One might think that this is trivial since the
sigma model for a second neighbor interaction is the same as for first neighbors. However if that would be the case
the correction would be $\lambda^2/J$ instead of the $\lambda^2/J^3$ as expected. We can see what happens by actually 
doing the calculation using:
\beqa
\vn_{k+p} &=& n_{k+\frac{p}{2}} + \frac{p}{2} \partial_\sigma \vn_{k+\frac{p}{2}}
+\frac{1}{2} \frac{p^2}{4} \ps^2 \vn_{k+\frac{p}{2}} +\frac{1}{6} \frac{p^3}{8} \ps^3 \vn_{k+{\frac{p}{2}}} 
+\frac{1}{24}\frac{p^4}{16} \ps^4\vn \ldots, 
\\ 
\vn_k \vn_{k+p} &=& 1 + \frac{p^2}{4} \vn\ps^2\vn - \frac{p^2}{4} \ps\vn\ps\vn -\frac{1}{48}p^4\ps\vn\ps^3\vn
 +\frac{p^4}{64} \ps^2\vn\ps^2\vn +\frac{p^4}{192}\vn\ps^4\vn + \ldots, \\
\sum_k  \vn_k \vn_{k+p} &=& \int d\sigma \left(1-\frac{p^2}{2} \ps\vn\ps\vn \ldots\right).
\eeqa
By replacing in (\ref{2loops}) we see that all terms up to (and including) cubic order in derivatives cancel. That is, the 
sigma model is higher order in derivatives, in this case four. Let us see how this affects the results. Since we want 
states with lowest energy for a given spin we expect $\vn$ to vary slowly. If the variations are of order one and 
the length of the chain is $J$ then we estimate that
\beq
\ps n \sim \frac{1}{J}.
\eeq
For the 1-loop calculation this gives
\beq
\gamma_1 \sim \lambda \int (\ps\vn)^2 \sim \lambda J \frac{1}{J^2} \sim \frac{\lambda}{J},
\eeq
as we obtained before. At two loops we have 
\beq
\gamma_2 \sim \lambda^2 \int \ps^2\vn\ps^2\vn \sim \frac{\lambda^2}{J^3},
\eeq
as we expect if the BMN limit is well defined. Furthermore, in the $n$-loop sigma model,      
all the terms having $2n-1$ derivatives or less should vanish. Those higher than $2n$ in derivatives are 
irrelevant in this limit. Therefore, for the full dilatation operator, including the Wess-Zumino term $S_{WZ}$, we 
should have the schematic expansion 
\beq
D = J + S_{WZ} + \lambda \int (\ps\vn)^2 + \lambda^2 \int (\ps\vn)^4 + \ldots + \lambda^n \int (\ps\vn)^{2n} \ldots
\eeq
This is just schematic since, at each order, one can have several terms with the same number of derivatives. This
is a peculiar low energy expansion where, each order in perturbation theory contributes at a certain order in 
momentum (or higher). It would be interesting to argue for this using supersymmetry perhaps as in~\cite{Mateos:2003de}.
It is also tempting to speculate that the absence of higher loop corrections to the $(\ps\vn)^2$ term is related
to the non-renormalization of the coupling constant in \N{4} SYM. 

 An important question is what is exactly the relation between the all-loop sigma model and the classical string action
in \adss{5}{5} (or in $\bbr{1} \times S^3$ for the case in hand). The Maldacena correspondence would indicate that they
are the same. However the sigma model we just wrote is first order in time derivatives and infinite order in spatial
derivatives. The Polyakov action is quadratic in both. It seems that the precise relation involves integrating
out the quadratic time derivatives and trading that for a non-local action in $\sigma$. In this first approach
we leave a more detailed analysis for future work.

%%%%%SECTION 4
\section{Conclusions}
\label{conclu}

In this paper we have shown that, as far as the one-loop anomalous dimension is concerned, the computation can be done 
using a sigma model in both, the field theory side and the string theory side. Fortunately, the sigma models agree. This 
shows that, at least in this particular sector of operators, agreement is guaranteed for any solution that can be found. 
It seems an interesting problem to generalize this to the full bosonic sigma model and 
perhaps to include the fermions.
In that case, from \cite{Beisert:2003jj,Beisert:2003yb} we expect to reproduce a 
limit of the Metsaev-Tseytlin action~\cite{Metsaev:1998it}. This might also shed light in the comparison of integrable
structures \cite{Bena:2003wd,Dolan:2003uh,Dolan:2004ps}.

 The most striking feature of the calculation is that the string appears in the field theory side. The time dependent
solution that we found is precisely the same as the rotating string. The expectation values of the spin is the
position in space time of the corresponding point of the string (after we discount the center of mass motion). It seems plausible
that this can allow us to derive a string dual for field theories for which they are not known. Of course there is no 
guarantee that the resulting string action corresponds to a critical superstring.  

We also discussed briefly the extension to higher loop order. For the solutions we analyzed, it does not seem too 
difficult to compute the two or higher loop corrections using the sigma model once the Hamiltonian for the spin 
chain is known. This is not so if one uses the Bethe ansatz. Of course now one would be 
more ambitious and try to match the sigma models without referring to particular solutions. It would be interesting
to see what picture emerges from there.

In particular it suggests that there is a precise formulation of the Maldacena correspondence which says that
the sigma model obtained from the all-loop spin chain Hamiltoninan is equivalent to the classical 
action of the string in \adss{5}{5}. At 1-loop the mapping is simply that they are both the same. At higher
loops the picture seems more complicated but we believe a careful analysis should indicate how to make
the formulation of the correspondence completely precise. By that I mean exactly what calculation one has to do 
in the field theory to recover the action of the string. We should perhaps clarify that this would be a formulation
of the weak version of the conjecture valid for large $N$. The strong version valid for any $N$ is simply
that the Hamiltonians and space of states completely agree. To go in that direction one has to include multitrace
operators and relate the ``splitting'' of a single trace operator into a double trace one with a 
string interaction \cite{Berenstein:2002jq,Beisert:2002ff}.
 
Many interesting field theories do not have scalar fields in the adjoint. Similar operators can be constructed
however as
\beq
\cO = \tr \nabla^{n_1} F \nabla^{n_2} F \nabla^{n_3} F \ldots  \nabla^{n_N} F,
\label{Fop}
\eeq
where $F$ is the field strength and we suppressed spatial indices. We see now that in each site an infinite 
number of derivatives (or states can appear). 
This is probably related to the emergence of a fifth dimension in the string dual, the radial dimension 
in \ads{} or the ``thickness'' in the QCD string.

Let us end with two speculations. What we have argued here is that, at least in a very restricted sector, the string action
emerges as the infrared limit of the dynamics of a discrete lattice. It is tempting then to argue that the usual, conformally
invariant action should be only the infrared fixed point of a more general, non-conformal model. There is going to be a regime
where the discrete model is applicable and another where the conformal theory is. However there has to be another, intermediate 
regime, where a continuum but non-conformal model is appropriate. Since the discrete description corresponds to 
the field theory whose description is valid for very large curvature and the conformal theory for small curvature, it is
natural to conjecture that a string in a background of curvature of order one (in string units) is described
by a non-conformal theory. How this can be true is not clear to us. 

The other speculation concerns the behavior of strings at finite temperature. Since we argued that the expectation value
of the spin gives the position of the string in space time, 
a high temperature disordered state of the spin chain looks in space time as a string whose shape resembles a random walk
rather that a continuum curve. We speculate that this can be a good description for strings above the Hagedorn temperature.

\section{Acknowledgments}
I am very grateful to Laura Urba for help in understanding the physics of the spin chain models. I am also very grateful
to Anton Ryzhov for interesting me in this subject and for close collaboration on related issues. 
 I am also indebted to A. Tseytlin for various comments on the first version of this paper. 
This work was supported in part by NSF through grants PHY-0331516, PHY99-73935 and DOE under grant DE-FG02-92ER40706.

\section*{NOTE ADDED}
After the first version of this paper appeared, in the works \cite{afterwork} some related issues
were clarified.


\begin{thebibliography}{99}        

%-----------------------------------------------------------------------------------

\bibitem{malda}
J.~Maldacena,
``The large $N$ limit of superconformal field theories and supergravity,''
Adv.\ Theor.\ Math.\ Phys.\  {\bf 2}, 231 (1998)
[Int.\ J.\ Theor.\ Phys.\  {\bf 38}, 1113 (1998)],
{\tt hep-th/9711200}, \\
%%CITATION = HEP-TH 9711200;%%
%\cite{Gubser:1998bc}
%\bibitem{Gubser:1998bc}
S.~S.~Gubser, I.~R.~Klebanov and A.~M.~Polyakov,
``Gauge theory correlators from non-critical string theory,''
Phys.\ Lett.\ B {\bf 428}, 105 (1998)
[arXiv:hep-th/9802109], \\
%%CITATION = HEP-TH 9802109;%%
%\cite{Witten:1998qj}
%\bibitem{Witten:1998qj}
E.~Witten,
``Anti-de Sitter space and holography,''
Adv.\ Theor.\ Math.\ Phys.\  {\bf 2}, 253 (1998)
[arXiv:hep-th/9802150], \\
%%CITATION = HEP-TH 9802150;%%
%\bibitem{magoo}
O.~Aharony, S.~S.~Gubser, J.~M.~Maldacena, H.~Ooguri and Y.~Oz,
``Large N field theories, string theory and gravity,''
Phys.\ Rept.\  {\bf 323}, 183 (2000)
[arXiv:hep-th/9905111].
%%CITATION = HEP-TH 9905111;%%

\bibitem{largeN}
G.'t Hooft, Nucl. Phys. {\bf B72} (1974) 461, 
G.'t Hooft, Nucl. Phys. {\bf B75} (1974) 461. 

%\cite{Berenstein:2002jq}
\bibitem{Berenstein:2002jq}
D.~Berenstein, J.~M.~Maldacena and H.~Nastase,
``Strings in flat space and pp waves from N = 4 super Yang Mills,''
JHEP {\bf 0204}, 013 (2002)
[arXiv:hep-th/0202021].
%%CITATION = HEP-TH 0202021;%%

%\cite{Gubser:2002tv}
\bibitem{Gubser:2002tv}
S.~S.~Gubser, I.~R.~Klebanov and A.~M.~Polyakov,
``A semi-classical limit of the gauge/string correspondence,''
Nucl.\ Phys.\ B {\bf 636}, 99 (2002)
[arXiv:hep-th/0204051].
%%CITATION = HEP-TH 0204051;%%

%\cite{Minahan:2002ve}
\bibitem{Minahan:2002ve}
J.~A.~Minahan and K.~Zarembo,
``The Bethe-ansatz for N = 4 super Yang-Mills,''
JHEP {\bf 0303} (2003) 013
[arXiv:hep-th/0212208].
%%CITATION = HEP-TH 0212208;%%


%\cite{Frolov:2003xy}
\bibitem{Frolov:2003xy}
S.~Frolov and A.~A.~Tseytlin,
``Rotating string solutions: AdS/CFT duality in non-supersymmetric  sectors,''
Phys.\ Lett.\ B {\bf 570}, 96 (2003)
[arXiv:hep-th/0306143].
%%CITATION = HEP-TH 0306143;%%

%\cite{Tseytlin:2003ii}
\bibitem{Tseytlin:2003ii}
A.~A.~Tseytlin,
``Spinning strings and AdS/CFT duality,''
arXiv:hep-th/0311139.
%%CITATION = HEP-TH 0311139;%%


%\cite{Braun:1998id}
\bibitem{Braun:1998id}
V.~M.~Braun, S.~E.~Derkachov and A.~N.~Manashov,
``Integrability of three-particle evolution equations in {QCD},''
Phys.\ Rev.\ Lett.\  {\bf 81}, 2020 (1998)
[arXiv:hep-ph/9805225], \\
%%CITATION = HEP-PH 9805225;%%
%\cite{Braun:1999te}
%\bibitem{Braun:1999te}
V.~M.~Braun, S.~E.~Derkachov, G.~P.~Korchemsky and A.~N.~Manashov,
``Baryon distribution amplitudes in {QCD},''
Nucl.\ Phys.\ B {\bf 553}, 355 (1999)
[arXiv:hep-ph/9902375],\\
%%CITATION = HEP-PH 9902375;%%
%\cite{Belitsky:1999ru}
%\bibitem{Belitsky:1999ru}
A.~V.~Belitsky,
``Integrability and WKB solution of twist-three evolution equations,''
Nucl.\ Phys.\ B {\bf 558}, 259 (1999)
[arXiv:hep-ph/9903512], \\
%%CITATION = HEP-PH 9903512;%%
%\cite{Belitsky:1999qh}
%\bibitem{Belitsky:1999qh}
A.~V.~Belitsky,
``Fine structure of spectrum of twist-three operators in {QCD},''
Phys.\ Lett.\ B {\bf 453}, 59 (1999)
[arXiv:hep-ph/9902361].
%%CITATION = HEP-PH 9902361;%%

%\cite{Beisert:2003xu}
\bibitem{Beisert:2003xu}
N.~Beisert, J.~A.~Minahan, M.~Staudacher and K.~Zarembo,
``Stringing spins and spinning strings,''
JHEP {\bf 0309}, 010 (2003)
[arXiv:hep-th/0306139].
%%CITATION = HEP-TH 0306139;%%

%\cite{Beisert:2003tq}
\bibitem{Beisert:2003tq}
N.~Beisert, C.~Kristjansen and M.~Staudacher,
``The dilatation operator of N = 4 super Yang-Mills theory,''
Nucl.\ Phys.\ B {\bf 664}, 131 (2003)
[arXiv:hep-th/0303060].
%%CITATION = HEP-TH 0303060;%%

%\cite{Beisert:2003ea}
\bibitem{Beisert:2003ea}
N.~Beisert, S.~Frolov, M.~Staudacher and A.~A.~Tseytlin,
``Precision spectroscopy of AdS/CFT,''
JHEP {\bf 0310}, 037 (2003)
[arXiv:hep-th/0308117].
%%CITATION = HEP-TH 0308117;%%

%\cite{Arutyunov:2003za}
\bibitem{Arutyunov:2003za}
G.~Arutyunov, J.~Russo and A.~A.~Tseytlin,
``Spinning strings in AdS(5) x S**5: New integrable system relations,''
arXiv:hep-th/0311004.
%%CITATION = HEP-TH 0311004;%%

%\cite{Arutyunov:2003uj}
\bibitem{Arutyunov:2003uj}
G.~Arutyunov, S.~Frolov, J.~Russo and A.~A.~Tseytlin,
``Spinning strings in AdS(5) x S**5 and integrable systems,''
Nucl.\ Phys.\ B {\bf 671}, 3 (2003)
[arXiv:hep-th/0307191].
%%CITATION = HEP-TH 0307191;%%

%\cite{Arutyunov:2003rg}
\bibitem{Arutyunov:2003rg}
G.~Arutyunov and M.~Staudacher,
``Matching higher conserved charges for strings and spins,''
JHEP {\bf 0403}, 004 (2004)
[arXiv:hep-th/0310182].
%%CITATION = HEP-TH 0310182;%%

\bibitem{Fradkin}
See \eg. chapter 5 in:\\ 
E. Fradkin, 
``Field theories of condensed matter systems,''
Addison-Wesley Publishing Company, Redwood City, CA, (1991).

\bibitem{Gradshteyn}
I.S. Gradshteyn, I.M. Ryzhik, ``Table of Integrals Series and Products'', Sixth edition,
Academic Press (2000), San Diego, CA, USA, London, UK.

\bibitem{wrapped}
U.~H.~Danielsson, A.~Guijosa and M.~Kruczenski,
``IIA/B, wound and wrapped,''
JHEP {\bf 0010}, 020 (2000)
[arXiv:hep-th/0009182], \\
%%CITATION = HEP-TH 0009182;%%
%\cite{Gomis:2000bd}
%\bibitem{Gomis:2000bd}
J.~Gomis and H.~Ooguri,
``Non-relativistic closed string theory,''
J.\ Math.\ Phys.\  {\bf 42}, 3127 (2001)
[arXiv:hep-th/0009181].
%%CITATION = HEP-TH 0009181;%%

%\cite{Gorsky:2003nq}
\bibitem{Gorsky:2003nq}
A.~Gorsky,
``Spin chains and gauge / string duality,''
arXiv:hep-th/0308182.
%%CITATION = HEP-TH 0308182;%

%\cite{Mikhailov:2003gq}
\bibitem{Mikhailov:2003gq}
A.~Mikhailov,
``Speeding strings,''
JHEP {\bf 0312}, 058 (2003)
[arXiv:hep-th/0311019].
%%CITATION = HEP-TH 0311019;%%

%\cite{Mateos:2003de}
\bibitem{Mateos:2003de}
D.~Mateos, T.~Mateos and P.~K.~Townsend,
``Supersymmetry of tensionless rotating strings in AdS(5) x S**5, and nearly-BPS operators,''
arXiv:hep-th/0309114.
%%CITATION = HEP-TH 0309114;%%

%\cite{Beisert:2003jj}
\bibitem{Beisert:2003jj}
N.~Beisert,
``The complete one-loop dilatation operator of N = 4 super Yang-Mills
theory,''
Nucl.\ Phys.\ B {\bf 676}, 3 (2004)
[arXiv:hep-th/0307015].
%%CITATION = HEP-TH 0307015;%%

%\cite{Beisert:2003yb}
\bibitem{Beisert:2003yb}
N.~Beisert and M.~Staudacher,
``The N = 4 SYM integrable super spin chain,''
Nucl.\ Phys.\ B {\bf 670}, 439 (2003)
[arXiv:hep-th/0307042].
%%CITATION = HEP-TH 0307042;%%

%\cite{Metsaev:1998it}
\bibitem{Metsaev:1998it}
R.~R.~Metsaev and A.~A.~Tseytlin,
``Type IIB superstring action in AdS(5) x S(5) background,''
Nucl.\ Phys.\ B {\bf 533}, 109 (1998)
[arXiv:hep-th/9805028].
%%CITATION = HEP-TH 9805028;%%

%\cite{Bena:2003wd}
\bibitem{Bena:2003wd}
I.~Bena, J.~Polchinski and R.~Roiban,
``Hidden symmetries of the AdS(5) x S**5 superstring,''
Phys.\ Rev.\ D {\bf 69}, 046002 (2004)
[arXiv:hep-th/0305116].
%%CITATION = HEP-TH 0305116;%%

%\cite{Dolan:2003uh}
\bibitem{Dolan:2003uh}
L.~Dolan, C.~R.~Nappi and E.~Witten,
``A relation between approaches to integrability in superconformal Yang-Mills
theory,''
JHEP {\bf 0310}, 017 (2003)
[arXiv:hep-th/0308089].
%%CITATION = HEP-TH 0308089;%%

%\cite{Dolan:2004ps}
\bibitem{Dolan:2004ps}
L.~Dolan, C.~R.~Nappi and E.~Witten,
``Yangian symmetry in D = 4 superconformal Yang-Mills theory,''
arXiv:hep-th/0401243.
%%CITATION = HEP-TH 0401243;%%

%\cite{Beisert:2002ff}
\bibitem{Beisert:2002ff}
N.~Beisert, C.~Kristjansen, J.~Plefka and M.~Staudacher,
``BMN gauge theory as a quantum mechanical system,''
Phys.\ Lett.\ B {\bf 558}, 229 (2003)
[arXiv:hep-th/0212269].
%%CITATION = HEP-TH 0212269;%%


\bibitem{afterwork}
%\cite{Mikhailov:2004qf}
%\bibitem{Mikhailov:2004qf}
A.~Mikhailov,
``Slow evolution of nearly-degenerate extremal surfaces,''
arXiv:hep-th/0402067,
%%CITATION = HEP-TH 0402067;%%
%\cite{Kazakov:2004qf}
%\bibitem{Kazakov:2004qf}
V.~A.~Kazakov, A.~Marshakov, J.~A.~Minahan and K.~Zarembo,
``Classical / quantum integrability in AdS/CFT,''
arXiv:hep-th/0402207,
%%CITATION = HEP-TH 0402207;%%
%\cite{Kruczenski:2004kw}
M.~Kruczenski, A.~V.~Ryzhov and A.~A.~Tseytlin,
``Large spin limit of AdS(5) x S**5 string theory and low energy expansion of
ferromagnetic spin chains,''
arXiv:hep-th/0403120.
%%CITATION = HEP-TH 0403120;%%
%\cite{Dimov:2004qv}
%\bibitem{Dimov:2004qv}
H.~Dimov and R.~C.~Rashkov,
``A note on spin chain / string duality,''
arXiv:hep-th/0403121,
%%CITATION = HEP-TH 0403121;%%
%\cite{Hernandez:2004uw}
R.~Hernandez and E.~Lopez,
``The SU(3) spin chain sigma model and string theory,''
arXiv:hep-th/0403139.
%%CITATION = HEP-TH 0403139;%%






\end{thebibliography}
\end{document}